\documentstyle[preprint,aps,psfig]{revtex} 
\begin{document}
\title{MOVING AVERAGES AND MARKETS INEFFICIENCY}
\author{R. Baviera$^{(1)}$, M. Pasquini$^{(2)}$, 
J. Raboanary$^{(3)}$ and M. Serva$^{(2,4)}$}
\address{$^{(1)}$Departement Finance \& Economie,
Groupe HEC, \\ Rue de la Liberation 1,
F-78351 Jouy-en-Josas, France}
\address{$^{(2)}$Istituto Nazionale 
di Fisica della Materia, I-67010 Coppito, L'Aquila, Italy}
\address{$^{(3)}$Institute Superieur Polytecnique de 
Madagascar, \\ Lot G III 32 Bis, Soamandrariny, Antananarivo,
 Madagascar}
\address{$^{(4)}$Dipartimento di Matematica,
Universit\`a dell'Aquila, 
I-67010 Coppito, L'Aquila, Italy}
\bigskip

\date{\today}

\maketitle

\begin{abstract}
We introduce a stochastic price model where, together with a random 
component, a moving average of logarithmic prices
contributes to the price formation.
Our model is tested against financial datasets, showing an 
extremely good agreement with them. 
It suggests how to construct trading strategies which
imply a capital growth rate larger than the
growth rate of the underlying asset, with also 
the effect of reducing the fluctuations.
These results are a clear evidence that some
hidden information is not fully integrated in price dynamics,
and therefore financial markets are partially inefficient.
In simple terms, we give a recipe for speculators to make money 
as long as only few investors follow it.
\end{abstract}

\section{Introduction}
\bigskip

Moving averages are probably the most popular and elemental analysis tool
in finance, widely known and applied both by professional 
and amateur traders.
Such a large favour is due to their simplicity and intuitive meaning,
which can help to understand the more or less hidden trend
of an evolving dataset, filtering some of the noise.

Moving averages are often studied 
and taken into account in modeling price dynamics both in financial
literature and in textbooks for market traders (technical analysis)
\cite{Pring,Meyers1,RL,OK,Meyers2,Taylor}.

In this paper we introduce a model for price dynamics where a moving average
plays a central role (in the following with 'price' we mean a more general
class of financial objects, including share quotes, indices, exchange rates,
and so on).  We focus our attention on the logarithm of price, not the
price itself, since the difference of two consecutive logarithmic prices
corresponds to price return, being a universal
measure of a price change, not affected by size factors. 

The moving average is therefore computed with logarithmic prices, 
but we do not simply look a time window in the past, giving
the same weight to each past quote. On the contrary, we take into
account every quote in the past, but its weight decays exponentially with 
time \cite{JBMS}. This choice is suggested by several evidences that 
correlations between price changes rapidly go to zero \cite{MS1,MS2}, 
at variance with absolute price changes (see for example \cite{PS1,PS2}).

In our price dynamics the difference between the logarithmic price
and the moving average at a given time linearly influences 
the future price, together with noise.
Let us stress that the moving average can either attract or repulse
the future price, being this a peculiar feature of the market considered.
In turns out that the logarithmic price plus 
its moving average evolves following a one-step Markov process.

We test our model analyzing three datasets:

a) Nasdaq index: 4032 daily closes from 1st Nov 1984 to 25th Sep 2000.

b) Italian Mibtel index: 1700 daily closes from 4th Gen 1994 to 2nd Oct 2000.

c) 1998 US Dollar - Deutsche Mark (USD-DM) bid exchange rate: 
1620861 high frequency data, that represent all worldwide 1998 bid quotes,
made available by Olsen \& Associated. 
The average time difference between two consecutive data is about $20$ seconds.

The results give a clear evidence that our model for price dynamics is
fully consistent with all the above datasets.

We also study the problem of finding an optimal strategy for
repeated trading operations in the context of our model, giving an analytic
solution for a realistic situation 
(small differences between logarithmic price and moving average).
In particular, we show that a simple suboptimal strategy 
implies a capital growth rate larger than the
growth rate of the underlying asset. 
This simple result highlights
a certain degree of market inefficiency.

Let us briefly sum up the contents of the paper.

In sect. II the model of price dynamics is introduced, and the exact
solution is derived.

In sect. III the model is tested against the three datasets, fixing 
the free parameters of the theory for each case.

In sect. IV strategies for repeated trades are investigated, in order
to find the optimal one.

In sect. V a suboptimal trading strategy is applied to the three datasets,
showing that markets hide some inefficiency.

In sect. VI some conclusions and some final remarks are presented.

\section{The model}
\bigskip

Price returns (i.e. the difference of two consecutive logarithmic prices)
are widely investigated in finance, since their dynamics does not depend on
size factors. For this reason we prefer to introduce a model for price dynamics
that focuses its attention on logarithmic price, instead of price itself.

Let us consider the price $S(t)$ of a financial object (an index,
an exchange rate, a share quote) at time $t$, and define the logarithmic price
as $x(t) \equiv \ln S(t)$. 
A moving average $\bar{x}(t)$ based on logarithmic prices, that takes into
account all the past with an exponentially time decaying weight,
can be written in the form

\begin{equation}
\bar{x}(t)= (1-\beta)\sum_{n=0}^{\infty}\beta^n x(t-n-1) \ \ .
\label{average}
\end{equation}
The moving average at time $t$ is computed only with past quotes with respect 
to $t$, and $x(t)$ is not included. This is not a relevant choice, since 
all the following could be reformulated in an equivalent way, 
including $x(t)$ in the moving average. 
The parameter $0<\beta<1$ controls the memory length:
for $\beta=0$ we have the shortest (one step) memory $\bar{x}(t)=x(t-1)$, while
when $\beta$ approaches $1$ the memory becomes infinity and flat (all the past
prices have the same weight).
More precisely, $\beta$ determines the typical past time scale 
$1-\ln 2/\ln\beta$ up to that we have a significant
contribution in the average (\ref{average}).

Let us suppose that future price linearly depends on the difference between 
the current price and the moving average, plus a certain degree of
randomness. In our model $x(t+1)$ is written as

\begin{equation}
x(t+1) = x(t) + \alpha \Big[ x(t)-\bar{x}(t) \Big] + \sigma\omega(t) \ \ ,
\label{model}
\end{equation}
where the $\omega (t)$ at varying $t$ are a set of independent identically
distributed random variables, with vanishing mean and unitary variance,
so that $\sigma \omega (t)$ is a random variable of variance $\sigma^2$.
Indeed, our analysis focuses on the problem 
of correlations with moving averages and we expect that results would 
not qualitatively depend on the shape of $\omega$ distribution,
which likely is a fat tails distribution \cite{Clark,Mandelbrot,MS3}.

The parameter $-1<\alpha<+1$ adjusts the impact of the difference
$x(t)-\bar{x}(t)$ over the future price. Notice that a positive $\alpha$ means
that the moving average is repulsive (the price most likely has a positive
change if $x(t)$ is larger then $\bar{x}(t)$), on the contrary
the moving average is attractive if $\alpha$ is negative. 
Also notice that the process loses all memory from the past
for $\alpha=0$, independently on $\beta$, becoming a pure random walk.
Let us stress that the Vasicek model \cite{Vasicek} is a particular case in our
more general scheme, corresponding to the peculiar choice $\beta=1$.

According to (\ref{model}), the return $r(t)$, defined as
$r(t) \equiv \ln \left( S(t+1) / S(t) \right)$, turns out to be
$r(t) = \alpha \Big[ x(t)-\bar{x}(t) \Big] + \sigma\omega(t)$.
Let us stress that the first contribution is measurable 
at time $t$ just before the price change and, therefore, 
is an available information for the trader. 

It is easy to check that the moving average (\ref{average})
satisfies

\begin{equation}
\bar{x}(t+1) = (1-\beta) x(t)+\beta \bar{x}(t) \ \ .
\label{aver-eq}
\end{equation}

Equations (\ref{model}) and (\ref{aver-eq}) define a two-component linear
Markov process. Such a process can be solved by a diagonalization
procedure. 
After having defined
$$
\tilde{\alpha}\equiv \frac {\alpha}{1-\beta} 
$$
and
$$
\epsilon \equiv \alpha + \beta
$$
which must satisfy $-1 <\epsilon <1$, we introduce
a new couple of variables, linear combinations of $x(t)$ and 
$\bar{x}(t)$ 
\begin{equation}
\begin{array}{ll}
y(t) \equiv x(t) - \bar{x}(t) \\
z(t) \equiv x(t) - \tilde{\alpha}\bar{x}(t) \ \ ,
\end{array}
\label{linear}
\end{equation}
being $y(t)$ simply the difference between the logarithmic price and its
moving average at time $t$.
The system turns out to be diagonal in terms of the new variables
$$
\begin{array}{ll}
y(t+1)= \epsilon y(t)+\sigma \omega(t) \\
z(t+1)= z(t)+\sigma \omega(t) \ \ ,
\end{array}
$$
and therefore can be easily solved, obtaining 
\begin{equation}
\begin{array}{ll}
y(t)= \sigma \sum_{n=0}^{t-1} \epsilon^n \omega(t-n-1)
+\epsilon^t y(0) \\
z(t)= \sigma \sum_{n=0}^{t-1} \omega(t-n-1)
+z(0) \ \ .
\end{array}
\label{sol-diag}
\end{equation}

The steady distribution $p_{\tilde{\sigma}}(y)$ of $y$
can be derived from equation (\ref{sol-diag}), once given
the $\omega$ distribution. The variance of $y$, $\tilde{\sigma}^2$,
can be directly  computed
$$
\tilde{\sigma}^2 \equiv \langle y^2 \rangle  =\frac{\sigma^2}{1-\epsilon^2} \ \ ,
$$
where $\langle.\rangle$ means average over the $\omega$ distribution.

The solution in terms of $x(t)$ and $\bar{x}(t)$ can be obtained
from (\ref{sol-diag}) by using the inverse transformation of 
(\ref{linear})
$$
\begin{array}{ll}
x(t) =x(0) + \frac{1-\beta}{1-\epsilon}
\left[ \sigma \sum_{n=0}^{t-1} (1-\tilde{\alpha}\epsilon^n) 
\omega(t-n-1)+ \tilde{\alpha}(1- \epsilon^t) 
\left[ x(0)-\bar{x}(0) \right] \right] \\  
\bar{x}(t)=\bar{x}(0) +\frac{1-\beta}{1-\epsilon} 
\left[ \sigma \sum_{n=0}^{t-1} (1-\epsilon^n) \omega(t-n-1)+ 
(1- \epsilon^t) \left[ x(0)-\bar{x}(0)\right] \right] \ \ . 
\end{array}
$$

\section{Data analysis}
\bigskip

In order to test our model against real financial data,
we need to find a measurable quantity that can be computed 
without specifying the $\omega$ distribution. 
For this reason, a direct comparison of the logarithmic prices
$x(t)$ with financial datasets cannot be of some help.
On the contrary the variance of any time window is independent
on the details of $\omega$ distribution.
It is, therefore, intuitive to consider $\rho(T)$, the $T$-days 
quadratic volatility scaled with $T$, defined as
$$
\rho(T) \equiv \frac{1}{T} \left\langle \left[ x(t+T)-x(t) \right]^2 \right\rangle \ \ .
$$
In the context of our model, the above quantity is
\begin{equation}
\rho(T) = \frac{\sigma^2}{T}
\frac{(1-\beta)^2}{(1-\epsilon)^2}
\Big[ T -
\frac{2\tilde{\alpha}
(1+\epsilon-\tilde{\alpha})}{1+\epsilon} \, \, \, 
\frac{1-\epsilon^T}{1-\epsilon} \Big] \, \, \, .
\label{Tvol}
\end{equation}

Notice that for $\alpha=0$ this quantity
would be a constant, while it increases with $T$ when $\alpha$ is positive 
(repulsive moving averages), and it decreases with $T$ when $\alpha$ is
negative (attractive moving averages). 
Notice also that the one-step ($T=1$) quadratic volatility is
$\rho(1)=\sigma^2 (1-\epsilon^2+\alpha^2)/(1-\epsilon^2)$,
which is always larger than $\sigma^2$.
In other terms, the one-step volatility is not simply $\sigma$,
the component due to the random variables, but it is systematically larger
due to the influence of the moving average.
In the long range ($T\to \infty$) $\rho(T)$ tends to
$\sigma^2 (1-\beta)^2 / (1-\epsilon)^2$, 
which is larger (smaller) then $\sigma^2$ for positive (negative) $\alpha$.

The expression (\ref{Tvol}) of $\rho(T)$ can be fitted with real data
varying the free parameters $\alpha$, $\beta$ and $\sigma$.
In fig. 1 $\rho(T)$ is plotted for three datasets:
Nasdaq index (fig. 1a), Mibtel index (fig. 1b) and 1998 USD-DM 
high frequency exchange rate (figs. 1c and 1d).

In the first two cases the agreement between data and expression (\ref{Tvol}) 
predicted by our model is excellent up to a
critical value of $T$, which depends on the size of the dataset.
For $T$ larger than the critical value insufficient statistics makes
the experimental values no longer significant.

In particular, we have obtained the following results:

a) Nasdaq index: $\alpha=(9.51 \pm 0.01) \cdot 10^{-2}$, 
$\beta=(2.49 \pm 0.01) \cdot 10^{-1}$ and  
$\sigma=(1.187 \pm 0.001) \cdot 10^{-2}$ in the range $1\le T\le 12$;

b) Mibtel index: $\alpha=(5.54 \pm 0.01) \cdot 10^{-2}$, 
$\beta=(3.15 \pm 0.01) \cdot 10^{-1}$ and  
$\sigma=(1.370 \pm 0.001) \cdot 10^{-2}$ in the range $1\le T\le 10$;

The two daily indices not only behave qualitatively the same,
but also exhibit similar parameters.
The main result is that they have a positive $\alpha$, 
which means that the moving average repulses the spot price.
Moreover, in both cases the typical memory length 
is about 2 time steps, which corresponds to a couple of trading days, 
and $\sigma$, which is nearly the daily market volatility,
turns out to be about $1\%$.
Let us stress that this is only the random component of daily volatility,
this last being larger because of the deterministic component contribution.

The 1998 USD-DM exchange rate case is more complex, since 
it exhibits a sort of transition phase. In fact
$\rho(T)$ turns out to be the union of two different curves
of type (\ref{Tvol}) (see fig. 1c):

c) 1998 USD-DM exchange rate: 

$\alpha=(-1.14 \pm 0.01) \cdot 10^{-1}$, 
$\beta=(1.57 \pm 0.01) \cdot 10^{-1}$ and  
$\sigma=(4.584 \pm 0.001) \cdot 10^{-4}$ in the range $1\le T\le 8$;

$\alpha=(-9.92 \pm 0.01) \cdot 10^{-1}$, 
$\beta=(9.00 \pm 0.01) \cdot 10^{-1}$ and  
$\sigma=(8.481 \pm 0.001) \cdot 10^{-4}$ in the range $8\le T\le 150$.

The agreement between data and the two fits is remarkable.
The transition point corresponds to $T=8$ (about 2.5 minutes), 
as shown in fig.1d where the range $1\le T \le 20$ is magnified.
This peculiar behaviour can be explained taking into account
that the price behaves like a quantum variable
on time scales from seconds to one or two minutes, according to a sort
of indeterminacy principle \cite{PS3}.
In fact, in \cite{PS3} it is shown that $T \rho(T)$ 
does not vanish in the $T \to 0$ limit, which means that 
two almost contemporary exchange rate bids not necessary 
coincide.
It is than reasonable to argue that 
the transition at $T=8$ in fig. 1c and fig. 1d corresponds
to the transition between quantum behaviour and classical
behaviour for the price.

In any case the USD-DM high frequency exchange rate has a
negative $\alpha$, which means that 
 the moving average attracts the spot price.
Moreover it exhibits a more consistent influence
of the moving average with respect to the indices, since $|\alpha|$
is about one order of magnitude larger, and the typical memory turns out
to be about 7-8 time steps (2.5 minutes).

\section{Efficiency}
\bigskip

A market is said efficient if prices reflect all the available information,
that is, a speculator cannot make a profit with probability one in the
long run. Evidence for a certain degree of inefficiency 
in financial markets has been recently found (see for example \cite{BPSVV}).

Our price dynamics model points out that there exists a contribution in price
formation of non random nature and, therefore, it is interesting to check
if the associated information is fully absorbed by the market or partially left
available for speculators.  
In this section we set up trading strategies based on our model, 
which will be applied to financial datasets in the next section: 
a performance much more successful than the implicit growth of the crude
data, given by a buy and hold strategy, 
will imply a clear evidence of inefficiency.

In order to do this check, let us consider a Kelly-like scheme of
repeated investments \cite{Kelly}: Kelly considered a long run 
of repeated random events, being fixed the probability of the outcome
of a single event and the fraction of capital bet at each time.
He proved that the capital growth rate is a non random quantity
in the limit of infinite events (in this limit the probability of have a growth
rate different from the typical value goes to zero). Then he was able
to maximize the capital growth rate with respect to the fraction of capital
invested, showing a linear relation between the rate itself and the Shannon
entropy of the random sequence.

In our case, given a capital $W(t)$,
$l_t$ is the fraction invested at time $t$ in an asset of price $S(t)$,
which can be modified at each time.
Since the price return $r(t)$ in our model depends only on the 
difference $\alpha \left(x(t)-\bar{x}(t)\right)= \alpha y(t)$ and on the
parameter $\sigma$, the same should be true for $l_t$, i.e. $l_t=l(\alpha y(t),
\sigma)$.  Then the capital $W(t)$ evolves according to
$$
W(t+1) = \Big[ 1-l_t+l_t \, \frac{S(t+1)}{S(t)} \Big]  W(t) \ \ ,
$$
and the associated growth rate is therefore
\begin{equation}
\lambda (\alpha,\beta,\sigma)\equiv
\lim_{T \to \infty } \frac{1}{T} \ln{\frac{W(T)}{W(0)}}
=\lim_{T \to \infty } \frac{1}{T} \sum_{t=0}^{T-1}  
\ln{\frac{W(t+1)}{W(t)}} =
\end{equation}
\begin{equation}
=\lim_{T \to \infty } \frac{1}{T} \sum_{t=0}^{T-1}  
\ln  \left(1-l_t+ l_t \, {\rm e}^ {\alpha y(t) +\sigma \omega(t)} \right) \ \ .
\label{rate}
\end{equation}

Our goal is to find the optimal $l^*(\alpha y(t), \sigma)$
which gives the maximum growth rate $\lambda^*(\alpha,\beta,\sigma)$ for a
given set $\alpha$, $\beta$ and $\sigma$.

Large number theory can be applied to right hand side of (\ref{rate})
since the variables in the sum are weakly correlated in time,
in fact, the $w(t)$ are uncorrelated, and the
$y(t)$ are related via the moving averages, whose correlations decay
exponentially by construction. 
As a consequence, the fraction of capital invested
no longer depends on $t$, and taking the maximum with
respect to $l$, one has
$$
\lambda^*(\alpha,\beta,\sigma)=
\int dy p_{\tilde{\sigma}} (y)
\mu(\alpha y, \sigma)  \ \ ,
$$
where $p_{\tilde{\sigma}} (y)$ is the steady distribution of variable $y$,
and
\begin{equation}
\mu (\alpha y, \sigma)=
\max_{l} \langle\ln (1-l+ l \, {\rm e}^ {\alpha y +\sigma \omega})\rangle \ \ ,
\label{mu}
\end{equation}
where the average $\langle.\rangle$ is performed over the $\omega$ distribution. 
The maximum in (\ref{mu}) is attained by $l=l^* (\alpha y, \sigma)$, 
according to the optimal strategy.  

Let us stress that $l$ must be restricted to $0\le l\le 1$ if
the $\omega$ distribution  is defined over all reals, to prevent 
that the logarithm argument in (\ref{mu}) becomes negative.

It is easy to show that for symmetric $\omega$ distributions
the following expressions hold:
$l^*(\alpha y, \sigma)=1-l^*(-\alpha y,\sigma)$ 
and $\mu (\alpha y,\sigma)=\alpha y +\mu (-\alpha y,\sigma)$.
Therefore, when $\alpha =0$ the maximum is attained at
$l^*(0,\sigma)=1/2$,  and one has: $\lambda (0,\beta,\sigma)=
\langle\ln \cosh(\sigma\omega/2)\rangle $, which is independent on $\beta$.

In a more general case, 
the result strictly depends on  $\omega$ distribution. 
Nevertheless, since in real financial data $\alpha y$ and
$\sigma$ are very small (about $1\%$ for daily data, and much
lower for intra-day data), it is reasonable to consider 
a Taylor expansion in $\alpha y$ and $\sigma$ up to second order.
The optimal $l^*(\alpha y,\sigma)$ turns out to be
\begin{equation}
l^*(\alpha y, \sigma)= \left\{
\begin{array}{lll}
0  & {\rm if} \ \ \alpha y \le -\frac{1}{2} \sigma^2 \\
\frac{1}{2}+\frac{\alpha y}{\sigma^2} \ \ & {\rm if} \ \
-\frac{1}{2} \sigma^2 <\alpha y < \frac{1}{2} \sigma^2 \\
1  & {\rm if} \ \ \alpha y \ge \frac{1}{2} \sigma^2 \\
\end{array}
\right. \ \ ,
\label{lstar}
\end{equation}
and, according to this result, one has
\begin{equation}
\mu(\alpha y, \sigma)= \left\{
\begin{array}{lll}
0  & {\rm if} \ \ \alpha y \le -\frac{1}{2} \sigma^2 \\
\frac{\sigma^2}{2} \left( \frac{1}{2}+\frac{\alpha y}{\sigma^2} \right)^2 \ \
& {\rm if} \ \ -\frac{1}{2} \sigma^2 <\alpha y < \frac{1}{2} \sigma^2 \\
\alpha y  & {\rm if} \ \ \alpha y \ge \frac{1}{2} \sigma^2 \\
\end{array}
\right. \ \ .
\label{mustar}
\end{equation}

In fig. 2 numerical exact results of $l^*(\alpha y, \sigma)$ are plotted
for a $\omega$ normal distribution and $\sigma=0.01$, compared with
approximation (\ref{lstar}). 
In fig. 3 the same comparison is performed for $\mu(\alpha y,\sigma)$,
plotting the numerical solution of (\ref{mu}) and the approximation
(\ref{mustar}).
Both graphs reveal the high quality of the approximated solution
in a context where variables assume realistic financial values.

\section{Forecasting}
\bigskip

Equation (\ref{lstar}) gives us a practical tool to construct 
an optimal trading strategy. Taking into consideration 
realistic values for $\alpha$ and $\sigma$, such as those found
in sect. III, one can realize that the $y$ interval with a non 
trivial $l^*$ is very small. As a consequence,
the optimal strategy (\ref{lstar}) is {\it de facto} almost equivalent to the
following more simple suboptimal strategy: invest all the capital ($l=1$)
whenever $\alpha y$ is positive, keep a neutral position ($l=0$) otherwise.

This simple trading strategy is applied to our three datasets, using
for parameters $\alpha$, $\beta$ and $\sigma$ the corresponding
values found in sect. III.
The resulting capital growth rates $\lambda(\alpha,\beta,\sigma)$
are compared with the intrinsic growth rates of crude data
$\lambda_{cd}$, which is obtained applying a buy and hold technique:

a) Nasdaq index: $\lambda(\alpha,\beta,\sigma) = 1.18 \cdot  10^{-3}$
and $\lambda_{cd} = 6.76 \cdot 10^{-4}$;

b) Mibtel index: $\lambda(\alpha,\beta,\sigma) = 9.60 \cdot 10^{-4}$
and $\lambda_{cd} = 6.80 \cdot 10^{-4}$;

c) USD-DM exchange rate: $\lambda(\alpha,\beta,\sigma) = 3.42 \cdot 10^{-5}$
and $\lambda_{cd} = -4.74 \cdot 10^{-8}$.

In the last case, we have used the parameters
resulting from the fit in the range $8\le T \le 150$.
For comparison
with the previous two cases concerning indices,
we have chosen to buy only dollars (the asset) against marks
(money),
while the opposite operation is not admitted.
The $\lambda_{cd}$ turns out to be negative since dollar
devalued during 1998.
The complete strategy should also allow 
to buy marks against dollars.
This strategy is about twice more successful than our,
but it loses its analogy with the one
we use in share markets.

In order to give further evidences of the good quality 
of the proposed strategy, 
we consider the logarithmic growth $\ln \left( W(t)/W(0) \right)$ 
of a capital invested according to our strategy
on the Nasdaq index. 
This simple strategy widely overcomes the index itself, as shown in fig. 4
where the logarithmic growth of the Nasdaq index is also plotted
for comparison. At the end of September 2000, while
the Nasdaq index is less than 16 times its initial value, 
the capital is become about 115 times.
In other terms, the applied strategy is able
to forecast the next quote with a probability of success greater than
$0.5$.

Another main feature of our strategy is revealed in fig. 4,
where the logarithmic scale allows to compare directly the 
relative fluctuations of the capital and of the Nasdaq index.
It is quite clear that the typical fluctuation of the capital is smaller
than that of the Nasdaq index. In particular, the impact 
of Nasdaq crashes on the capital is smoothened out.

The above results clearly prove a certain degree
of inefficiency in the markets under consideration, since there exists
at least a deterministic procedure to make a sure profit
on a long run.
A very interesting question is to determine how much of this
information, not integrated by markets, keeps available
for speculators once transactions cost are taken into account.
Roughly speaking, we find a daily growth rate of about $0.1\%$
with the strategies involving indices where 
trading frequency is less than an operation per day.
It follows that a capital of about ten thousands of 
dollars should produce profit in case of fixed transaction costs of
the order of few dollars per trade. 
These conditions are quite realistic both in
american and european markets.

\section{Conclusions}
\bigskip

Moving average plays an important role in price dynamics.
Future price is influenced by the current difference between 
logarithmic price and its moving average. This difference
tends both to reduce and to enlarge,
according to the examined market.
The first open question is: this feature is a strict peculiarity
of the market, or does it depends on the time frequency of data?
In our cases, both daily datasets exhibit a repulsive moving average,
at difference with USD-DM high frequency dataset.
The answer could be that at different time scales, different 
moving average effects are active, possibly attractive on short
scales and repulsive on larger scales. The problem deserves further
investigations.

Another interesting point is to investigate
if the moving average action is generated by 
a self-organized mechanism of traders reactions \cite{Zhang,Palmer}. 
In other words, is it possible that traders, taking into account
informations given by moving averages, 
make collectively induced financial choices producing,
as a result, the observed phenomena?

In our price dynamics model a random component
is also present,
which we do not have deeply investigated, being this
out of the scope of this paper. 
Nevertheless, our picture could help
to determine the exact shape of the noise, 
since, in principle, we are now able to filter the 
deterministic contribution.

Finally, in the light of our model, we have found a trading strategy
that widely overcomes the intrinsic performance of the examined
datasets. In this way, we have given a clear evidence
that some inefficiency is present in financial markets.
Next challenge is to find out if part of this
inefficiency survives when applied by a real speculator,
which lives in a real financial world, where transaction costs, 
unfortunately, are not omitted.

\bigskip
\bigskip
\centerline{\bf Acknowledgments}
We thank Dietrich Stauffer
for many useful discussions
and for a critical reading of the manuscript.

\begin{figure}
\mbox{\psfig{file=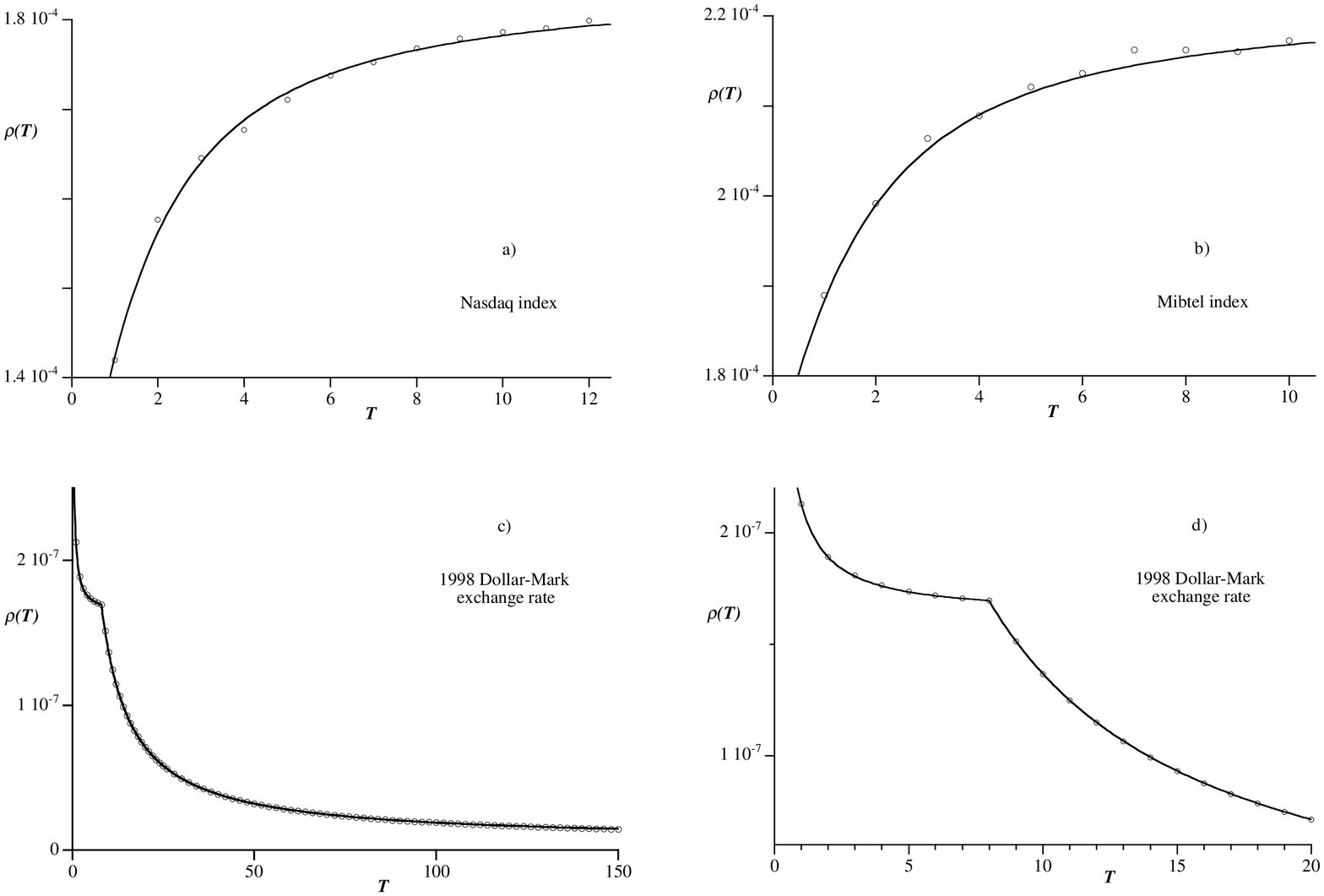,width=16cm}}
\caption{
$T$-days quadratic volatility scaled with $T$, $\rho(T)$ (circles), 
as a function of $T$, for:
a) Nasdaq index; b) Mibtel index; c) 1998 USD-DM high frequency exchange rate;
d) magnification of the previous case in the range
$1\le T \le 20$.
The continuous line represents the best fit of expression (\ref{Tvol}), 
which gives: \\
a) Nasdaq index: $\alpha=(9.51 \pm 0.01)  \cdot 10^{-2}$, 
$\beta=(2.49 \pm 0.01) \cdot 10^{-1}$ and 
$\sigma=(1.187 \pm 0.001) \cdot 10^{-2}$; \\
b) Mibtel index: $\alpha=(5.54 \pm 0.01) \cdot 10^{-2}$, 
$\beta=(3.15 \pm 0.01) \cdot 10^{-1}$ and 
$\sigma=(1.370 \pm 0.001) \cdot 10^{-2}$; \\
c-d) 1998 USD-DM exchange rate: 
$\alpha=(-1.14 \pm 0.01) \cdot 10^{-1}$, 
$\beta=(1.57 \pm 0.01) \cdot 10^{-1}$ and  
$\sigma=(4.584 \pm 0.001) \cdot 10^{-4}$ in the range $1\le T\le 8$;
$\alpha=(-9.92 \pm 0.01) \cdot 10^{-1}$, 
$\beta=(9.00 \pm 0.01) \cdot 10^{-1}$ and  
$\sigma=(8.481 \pm 0.001) \cdot 10^{-4}$ in the range $8\le T\le 150$.
}
\end{figure}

\begin{figure}
\mbox{\psfig{file=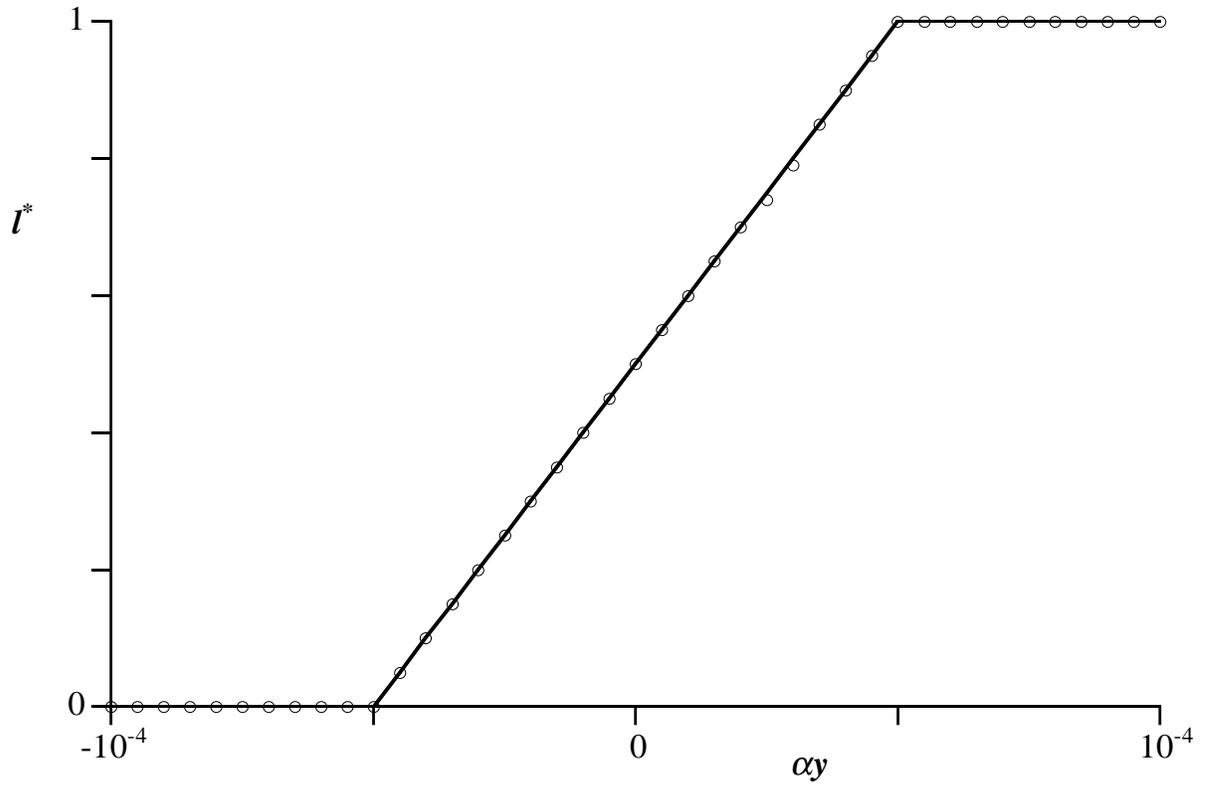,width=16cm}}
\caption{
$l^*(\alpha y,\sigma)$, the value which maximizes the
expression (\ref{mu}), as a function of $\alpha y$ for a $\omega$ normal distribution 
and $\sigma=0.01$: numerical solution (circles) compared with
approximation (\ref{lstar}) (continuous line).
}
\end{figure}

\begin{figure}
\mbox{\psfig{file=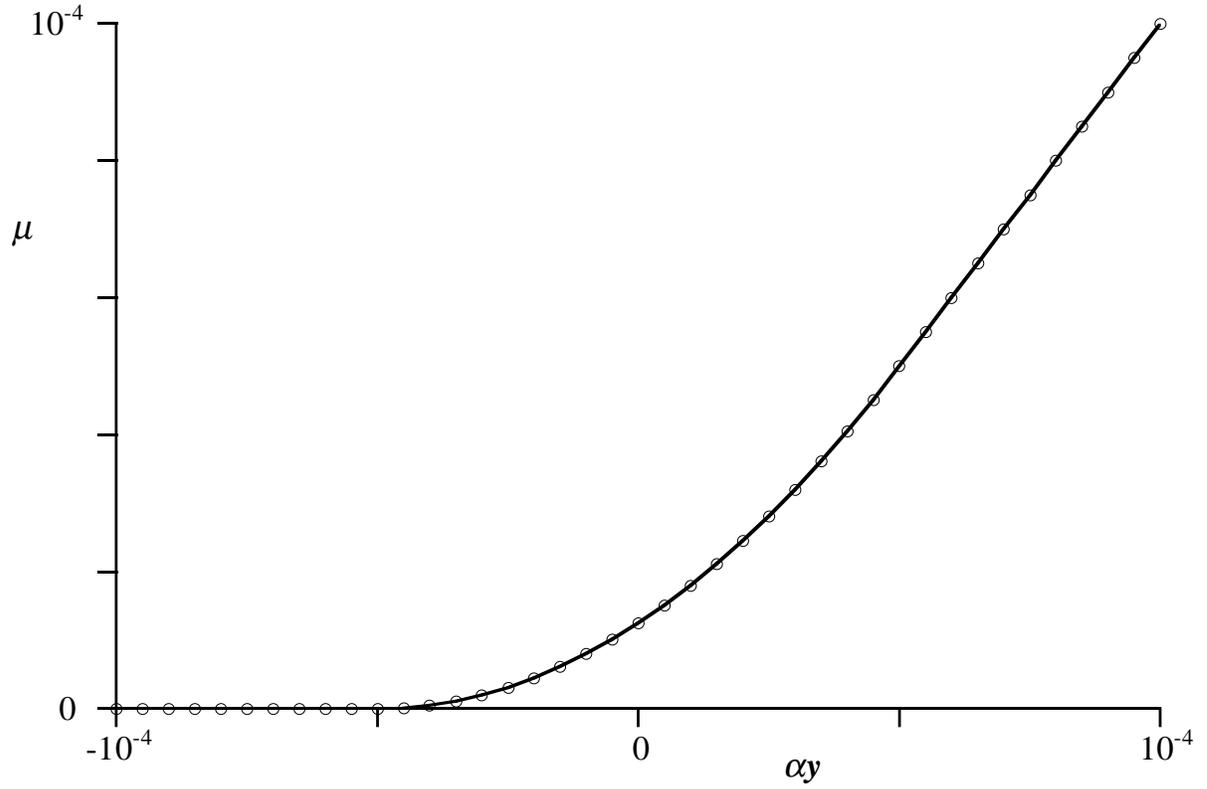,width=16cm}}
\caption{
$\mu(\alpha y,\sigma)$ (\ref{mu}) as a function of $\alpha y$ for a $\omega$ normal 
distribution and $\sigma=0.01$: numerical solution (circles) compared
with approximated solution (\ref{mustar}) (continuous line).
}
\end{figure}

\begin{figure}
\mbox{\psfig{file=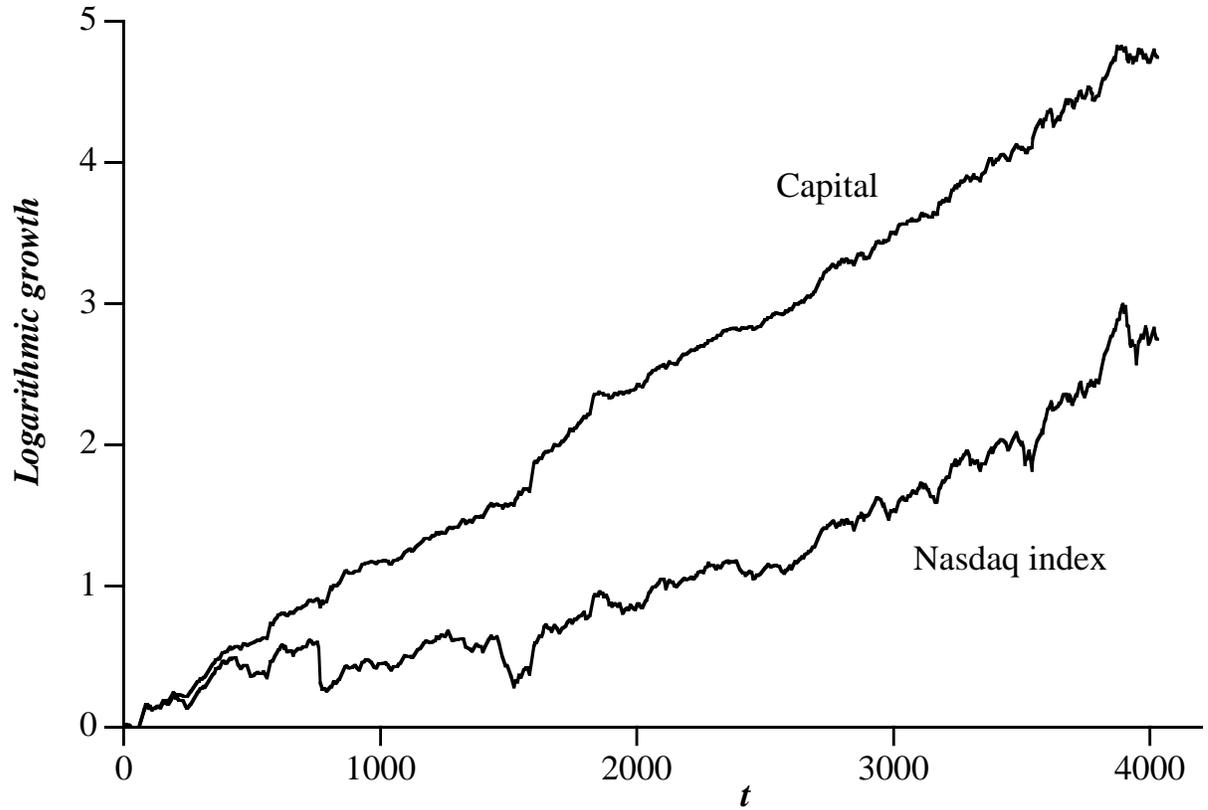,width=16cm}}
\caption{
Logarithmic growth $\ln \left(W(t)/W(0)\right)$ 
of a capital $W(t)$ invested on the Nasdaq index as a function of time $t$, 
compared with the logarithmic growth of the index itself.
}
\end{figure}

\end{document}